\documentclass[a4paper,11pt]{article}
\usepackage{pos}
\usepackage{xspace}

\newcommand{\beq}{\begin{equation}}
\newcommand{\eeq}{\end{equation}}
\newcommand{\bea}{\begin{eqnarray}}
\newcommand{\eea}{\end{eqnarray}}
\newcommand{\qedl}{${\rm QED}_L$ }

\newcommand{\VTC}{V_{\rm TC}}
\newcommand{\VCTC}{V_{\overline{\raisebox{2.3 mm}{}\mathrm{TC}}}}
\newcommand{\KCTC}{K_{\overline{\raisebox{2.3 mm}{}\mathrm{TC}}}}

\newcommand{\MCTC}{M_{\overline{\raisebox{2.3 mm}{}\mathrm{TC}}}}
\newcommand{\dCTC}{\delta^{\overline{\raisebox{2.3 mm}{}\mathrm{TC}}}}

\newcommand{\ba}{\begin{eqnarray}}
\newcommand{\ea}{\end{eqnarray}}

\newcommand{\be}{\begin{equation}}
\newcommand{\ee}{\end{equation}}

\title{Coulomb corrections to pi-pi scattering}

\author[a]{Norman Christ}
\author[b,c,d,e]{Xu Feng}
\author*[a]{Joseph Karpie}
\author[a]{Tuan Nguyen}

\emailAdd{jmk2289@columbia.edu}

\affiliation[a]{Physics Department, Columbia University, \\
New York City, New York 10027, USA}
\affiliation[b]{Collaborative Innovation Center of Quantum Matter, \\
Beijing 100871, China}
\affiliation[c]{Center for High Energy Physics, Peking University, \\
Beijing 100871, China}
\affiliation[d]{School of Physics, Peking University, \\
Beijing 100871, China}
\affiliation[e]{State Key Laboratory of Nuclear Physics and Technology, Peking University, \\
Beijing 100871, China}

\abstract{The relationship between finite volume multi-hadron energy levels and matrix elements and two particle scattering phase shifts and decays is well known, but the inclusion of long range interactions such as QED is non-trivial. Inclusion of QED is an important systematic error correction to $K\to\pi\pi$ decays. In this talk, we present a method of including a truncated, finite-range Coulomb interaction in a finite-volume lattice QCD calculation. We show how the omission caused by the truncation can be restored by an infinite-volume analytic calculation so that the final result contains no power-law finite-volume errors beyond those usually present in Luscher’s finite-volume phase shift determination. This approach allows us to calculate the QED-corrected infinite-volume phase shift for $\pi\pi$ scattering in Coulomb gauge, a necessary ingredient to $K\to\pi\pi$, while neglecting the transverse radiation for now.}

\FullConference{%
 The 38th International Symposium on Lattice Field Theory, LATTICE2021
  26th-30th July, 2021
  Zoom/Gather@Massachusetts Institute of Technology
}

\note{For the RBC/UKQCD collaboration.}

\begin{document}
\maketitle

\section{Introduction}
Neutral Kaon decay into pions is fundamental to the understanding of $CP$ violation from the Weak interaction. In $K_L\to \pi\pi$ decays, indirect $CP$ violation is generated by mixing between $K^0$ and $\bar{K^0}$ causing a small $CP$ even component to $K_L$ parameterized by $\epsilon$. Direct $CP$ violation, parameterized by $\epsilon'$, comes from the leading $CP$ odd component of $K_L$ and the non-zero phases of the CKM matrix. Experimental measurements have shown that the direct violation parameter is significantly smaller than the indirect parameter with ${\rm Re}(\epsilon'/\epsilon) = 16.6(2.3) \times 10^{-4}$~\cite{Batley:2002gn, PhysRevD.83.092001}. Recently, a lattice QCD calculation has been performed to obtain a standard model prediction ${\rm Re}(\epsilon'/\epsilon) = 21.7(2.6)(6.2)(5.0) \times 10^{-4}$~\cite{Abbott:2020hxn}. The reported errors are first the statistical error, second the systematic error, and third the error from neglecting the isospin violating effects of electromagnetism and the light quark mass different $m_u-m_d$. Since the ratio ${\rm Re}(\epsilon'/\epsilon)$ is so small, it can be sensitive to new physics and a precision standard model estimation is of great interest. 

The largest individual source of error in the standard model calculation is the neglect of isospin breaking effects. Typically, such effects are of $O(1\%)$, but in the calculation of $\epsilon'$, they are exacerbated by the $\Delta I=1/2$ rule. With perfect isospin, $\epsilon'$ is given by
\beq
\epsilon' = \frac{i e^{\delta_2 -\delta_0}}{\sqrt 2} \frac{{\rm Re} A_2}{{\rm Re} A_0} \left(\frac{{\rm Im} A_2}{{\rm Re} A_2} - \frac{{\rm Im} A_0}{{\rm Re} A_0} \right)
\eeq
where $A_{0,2}$ are the decay amplitudes for $K^0$ to go into $I=0,2$ $\pi\pi$ states and $\delta_{0,2}$ are the $I=0,2$ $\pi\pi$ scattering phase shifts. The $\Delta I=1/2$ rule is that $A_2$ is suppressed, relative to $A_0$, by a factor of 22. The $O(1\%)$ mixing of the isospin states at $O(\alpha_{EM})$ or $O(m_u-m_d)$, will be amplified to a $O(20\%)$ correction when propagated to the calculation of ${\rm Re}(\epsilon'/\epsilon)$. Chiral perturbation theory has been used~\cite{Cirigliano:1999ie, Cirigliano:1999hj, Cirigliano:2000zw, Wolfe:2000rf, Cirigliano:2003gt, Cirigliano:2019cpi} to study the size of the electromagnetic effects, but since the corrections are so large, a direct ab initio lattice QCD calculation is desirable. How to handle this calculation has been studied before in~\cite{Christ:2017pze,Cai:2018why}.

There exist a number of complications which must be understood before embarking on a full lattice QCD calculation of the isospin breaking contributions to $\epsilon'$. The first which we will attempt to handle is caused by numerical lattice QCD's necessary finite volume. The calculation of $K\to \pi\pi$ decays utilizes the methods of L\"uscher~\cite{Luscher:1990ux} to relate finite volume multi-particle energy states to infinite volume scattering phase shifts and Lellouch-L\"uscher~\cite{Lellouch:2000pv} to relate finite volume correlation functions to infinite volume decay amplitudes. These methods require the assumption that the interactions are exponential localized which is violated by the long distance electromagnetic interactions. How to avoid this complication will be the subject of this talk. A secondary complication is that with isospin symmetry the two $\pi\pi$ final states are independent. With isospin breaking effects included, determining the final scattering state is a coupled two-channel problem which complicates the Lellouch-L\"uscher approach. Finally, there will exist well IR singularities from near-degenerate final states of (the desired) $\pi\pi$ states and those with two pions and any number of soft photons. These expand the previously mentioned difficulty to not just include $I=0$ and $I=2$ $\pi\pi$ state mixing, but also more complicated three particle channels. 

Before considering the complicated full problem of $K^0$ decays, we will focus on the first problem of including electromagnetic effects into the standard finite volume techniques. We begin with considering $\pi^+\pi^+$ scattering, which lacks the isospin mixing multichannel state, and study how to make QED compatible with L\"uscher's approach. This process is done by dividing the problem into more manageable parts. First, we select to work with QED in the Coulomb gauge. Since within a lattice QCD calculation, it is natural to fix the frame into the Kaon's rest frame, the choice of a non-covariant gauge will not introduce later difficulties. The advantage of the Coulomb gauge is electromagnetic interactions are separated into an instantaneous Coulomb interaction and transverse photon radiation. Both of these pieces can be handled independently in a lattice QCD calculation. In this talk, we will neglect the transverse radiation piece. 

The remaining Coulomb interaction will be truncated into two regions. The short distance region will be such that L\"uscher's finite volume approach will be perfectly justified. The long distance region will be such that it can be analyzed analytically in infinite volume outside a lattice QCD calculation. A similar division could be performed for the transverse radiation. Photons above a certain energy will be allowed into the finite volume lattice QCD calculation, while those below said energy will be handled in infinite volume with the standard Bloch-Nordsiek methods. It should be noted that this is an alternative approach to including QED into phase shift calculations from lattice QCD than was presented in~\cite{Beane:2014qha, NPLQCD:2020ozd}. They use the conventional \qedl approach, which modifies the Coulomb interaction to be periodic in a finite volume. This is done by introducing new power law finite volume errors which must be corrected for. Only the leading $1/L$ corrections are universal and higher powers must be removed by studying the volume dependence. The truncated Coulomb approach will introduce errors in the lattice QCD calculation based on the truncation radius, but these should be corrected by the long distance analytical calculation. 

The Coulomb interaction will be dividing into the two pieces
\begin{eqnarray}
\VTC &=&  \frac{1}{2} \int d^3 r d^3 r^\prime \rho(\vec r)\,\frac{\theta\bigl(R-|\vec r-\vec r\,'|\bigr)}{4\pi|\vec r - \vec r\,^\prime|}\rho(\vec r\,^\prime) 
\label{eq:VTCop} \\
\VCTC &=&  \frac{1}{2} \int d^3 r d^3 r^\prime \rho(\vec r)\,\frac{\theta\bigl(|\vec r-\vec r\,'|-R\bigr)}{4\pi|\vec r - \vec r\,^\prime|}\rho(\vec r\,^\prime)\,. \label{eq:VCTCop}
\end{eqnarray}
$\VTC$ is the truncated Coulomb potential which governs the interactions with separations $|\vec r - \vec r\,^\prime| <R$ and will be discussed in Sec.~\ref{sec:sd}. $\VCTC$ is the complement of the truncated Coulomb potential with governs interactions with separations $|\vec r - \vec r\,^\prime| >R$ and will be discussed in Sec.~\ref{sec:ld}

\section{Numerical Treatment for Short Distance}~\label{sec:sd}
The short distance Coulomb interaction $\VTC$ clearly can satisfy the conditions for the standard L\'uscher finite volume quantization condition. In that framework, the interactions' strengthes must be exponentially suppressed at distances comparable to the volume, or more specifically half the volume, for there to be no power law finite volume corrections to the derived phase shift~\cite{Luscher:1990ux}. Given $R<L/2$, this condition is naturally satisfied, even if $R$ is close to $L/2$, though a numerical study is necessary to demonstrate how well that limit holds. 

As in calculations with \qedl, $\VTC$ can be added perturbatively or non-perturbatively. With a non-perturbative insertion, the quantization condition can be applied as prescribed by L\"uscher~\cite{Luscher:1990ux}. 
\beq
\delta_0(p) + \phi(q) = n\pi\,,
\eeq
where the pion's momentum is $p=\sqrt{(E/2)^2 -m^2}$ for a given two particle energy $E$ in a periodic finite volume of size $L$ and $q=\frac{Lp}{2\pi}$. The phase $\phi$ is defined by
\beq
\tan \phi(q) = -\frac{\pi^{3/2}q}{Z_{00}(1,q)}
\eeq
where $Z_{00}$ is the L\"uscher Zeta function. This quantization condition will give the $l=0$ phase shift up to exponential finite volume corrections and power law corrections for ignoring contributions from $l=4$ and higher states. The latter corrections could in principle be determined. 

One could also implement $\VTC$ in perturbation theory. In this case would would calculate the energy shifts for the perturbative expansion $E = E^{(0)} + \alpha E^{(1)} + \dots$. The first order correction can be determined from a lattice QCD calculation in a periodic finite volume of size $L$. Given a suitable 2 pion interpolating field $O_{\pi\pi}(t)$, the ground state energy shift would be given by the ratio of correlation functions
\beq\label{eq:latt_pert}
E^{(1)} = \frac{\langle O_{\pi\pi}(t_f) \frac12 \int d^3r_1 d^3r_2 \rho(r_2,t_V) \VTC(|r_1 - r_1|_L) \rho(r_1,t_V) O_{\pi\pi}(t_i) \rangle}{\langle O_{\pi\pi}(t_f)O_{\pi\pi}(t_i) \rangle}
\eeq
if the time separations $t_f - t_V$ and $t_V-t_i$ are sufficiently large that excited states are suppressed. $\rho(r,t)$ is an appropriate charge density operator and the argument of $\VTC$ is meant to be the shortest periodic distance between $r_1$ and $r_2$,
\beq
|r_1 - r_1|_L = \left\{ \sum_{i=1}^3 \bigg( \min\big[|(r_2)_i - (r_1)_i|, L - |(r_2)_i - (r_1)_i)|  \big] \bigg)^2 \right\}^{1/2} \,.
\eeq
If one wishes to determine the energy shift for the lowest excited state, then the time dependences of the same ratio must be studied, using standard techniques in lattice QCD. After obtaining the energy shifts, one can perform the same perturbative expansion of the phase shift, $\delta_0 = \delta_0^{(0)} + \alpha \delta_0^{(1)} + \dots$. By isolating the $O(\alpha)$ terms, one can arrive at the first order correction to the phase shift
\beq\label{eq:phase_pert}
\delta_0^{(1)}(p^{(0)}) =  - \left\{ \frac{d\delta_0(p)}{dp} + \frac{d\phi(q)}{dq} \frac{L}{2\pi} \right\}_{p=p^{(0)}} \frac{E^{(0)}}{4p^{(0)}} E^{(1)}\,,
\eeq
given $p^{(0)} = \sqrt{(E^{(0)}/2)^2 -m^2}$. Through Eq.~\eqref{eq:phase_pert} and similar derivations for higher orders, one can obtain the phase shift from QCD and the truncated Coulomb interaction order by order.

A final consideration is the renormalization of the pion's mass from the Coulomb interaction. One wants the pion mass in the $O(\alpha)$ calculation to have the same physical value as in the chargeless $O(\alpha^0)$ case. First, one must find the single pion energy shift caused by the Coulomb interaction in identical fashion to Eq.~\eqref{eq:latt_pert}, but with a single pion interpolating field. Then, one determines the quark mass shift $\alpha m^{(1)}$ which compensates for the Coulomb interaction in order to keep the physical pion mass the same. A scalar quark bilinear interaction with coefficient $\alpha m^{(1)}$ can then be added to the $\VTC$ operator in Eq.~\eqref{eq:latt_pert}. It is the resulting energy shift from this modified interaction which will be used in Eq.~\eqref{eq:phase_pert} to determine the perturbative change in the phase shift, while keeping the physical pion masses fixed.

\section{Analytic Treatment for Long Distance}~\label{sec:ld}
The long distance nature of the Coulomb interaction is a significant difficulty for implementations in a finite volume. The corrections to the phase shift from purely the long distance portion of the Coulomb interaction, given by $\VCTC$, can be calculated analytically in an infinite volume and approximating the pions as having a purely local interaction such as a $(\phi^\dagger\phi)^2$ theory. This correction can be determined in terms of the phase shift $\delta_l$ without QED interactions and the electromagnetic form factor of the pion. The correction are calculable up to terms which are exponentially suppressed in $R$ as long as the energy is below the four pion threshold.

\begin{figure}
\centering
\includegraphics[width=1.0\linewidth]{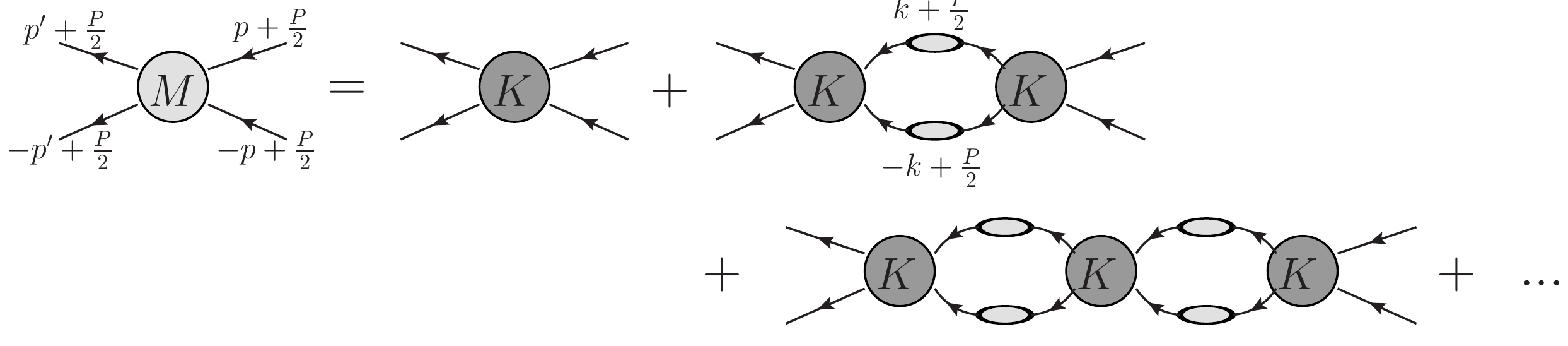}
\caption{A diagram describing the Lippman-Schwinger equation. The scattering amplitude $M$ in terms of products of a 2-particle irreducible kernel $K$ multiplying by pairs of dressed pion propagators, shown by the lines with a lightly shaded bubble. }
\label{fig:LS}
\end{figure}

The analytic calculation begins with a relativistic Lippman-Schwinger equation, with the scattering amplitude given by a series of products of a four pion, two particular irreducible, scattering kernels joined by intermediate pairs of pion propagators.  This series is shown diagramatically in Fig.~\ref{fig:LS}. The four pion scattering kernel will be assumed to be at all order in the $(\phi^\dagger\phi)^2$ interaction. If the scale $R$ is significantly longer than the relevant QCD distance scale, such as $\Lambda_{\rm QCD}^{-1}$, then we can assume that the results of this analytic calculation can be interpreted universally. The on shell center-of-mass scattering amplitude $M_l(E)$ is defined by
\beq
iM_l(E) \delta_{l'l} \delta_{m'm} = \frac{1}{4\pi}\int\!\!\int d\Omega_{\hat p'} d\Omega_{\hat p}\,
                                                      Y^*_{l'm'}(\hat p') M\bigl((\vec p\,',\omega_p),(-\vec p\,',\omega_p),(\vec p,\omega_p),(-\vec p,\omega_p)\bigr) Y_{lm}(\hat p).
\label{eq:PW-projection}
\eeq
where $\vec p$ and $\vec p'$ are 3-momenta of magnitude $p$ whose directions are integrated, $\omega_p$ are the related pion energies, and for on shell momenta $M(p_4, p_3, p_2, p_1)$ is the obtained from the connected time-ordered product
\begin{eqnarray}
&& \prod_{i=1}^4 \left\{\frac{p_i^2+m^2}{i}\right\}\prod_{i=1}^4 \int \left\{\int d^4 x_i \right\} 
e^{i(-p_4x_4-p_3x_3+p_2x_2+p_1x_1)}
                  \bigl\langle\phi(x_4)\phi(x_3)\phi^\dagger(x_2)\phi^\dagger(x_1)\bigr\rangle_\mathrm{conn} \nonumber \\
&& \hskip 1.5 in = (2\pi)^4\delta^4(p_4+p_3-p_2-p_1) M(p_4,p_3,p_2,p_1)\,.
\label{eq:time-ordered}
\end{eqnarray}
When below the four pion threshold, using unitarity, the scattering amplitude defines the phase shifts as
\beq
M_l(E) = 32\pi\frac{\omega_p}{p}\frac{e^{2i\delta_l}-1}{2i}\,.
\label{eq:phase-shift}
\eeq
When calculating this amplitude, it will prove beneficial to actually determine the Euclidean space analogue and then analytically continue it into Minkowski space. The validity of this analytic continuation is discussed in full detail in~\cite{Christ:2021guf}.

\begin{figure}
\centering
\includegraphics[width=0.6\linewidth]{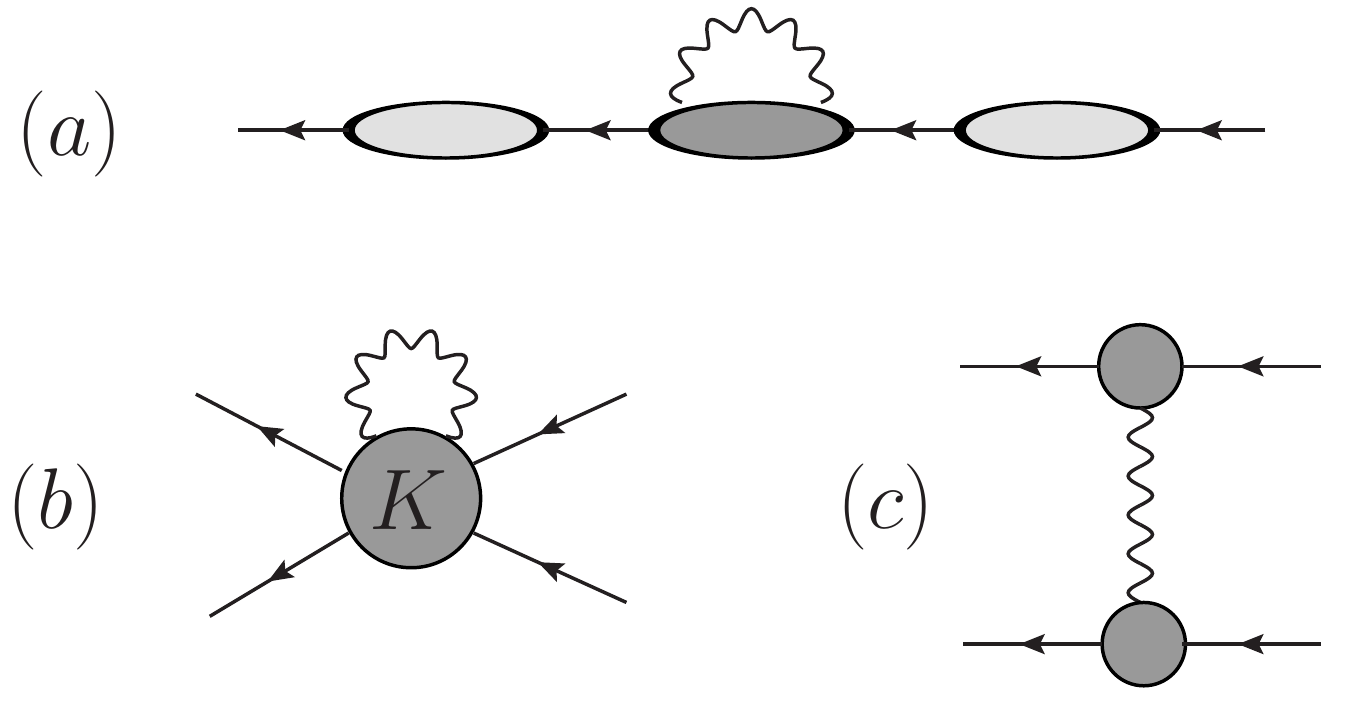}
\caption{The possible subdiagrams demonstrating how $\VCTC$ can be included into the series shown in Fig.~\ref{fig:LS}. Diagram $(a)$ is the self energy correction for one of the dressed pion propagators. Diagram $(b)$ is a modified two-particle irreducible kernel which cannot be broken into two pieces by removing the $\VCTC$ interaction's photon line. Diagram $(c)$ is a modified two-particle irreducible kernel which can be broken into two pieces when the photon line is cut. It is called the exchange diagram and is the only class of diagrams which can have power law corrections in $R^{-1}$.}
\label{fig:AddCTC}
\end{figure}

There are three classes of diagrams of how the $\VCTC$ interaction can be included at first order in perturbation theory, shown in Fig.~\ref{fig:AddCTC}. The first case $(a)$ is a self energy diagram in one of the pion propagators connected the 2 particle irreducible kernels. The second and third cases $(b)$ and $(c)$ are when the interaction occurs within the 2 particle irreducible kernel. The interaction occurs between electromagnetic currents at positions $(z_1,t)$ and $(z_2,t)$ which can be shown graphically with a wavy ``photon'' line in this new $O(\alpha)$ kernel. If cutting this line separates the new kernel into two distinct parts, then it is of type $(c)$, otherwise of type $(b)$. It is only diagrams of type $(c)$, where the Coulomb interaction is between two well separated pions, which are not exponentially suppressed in $R$ when the energy is below the four pion threshold. In the other two types of diagrams, at least two pions must travel the long space-like distance $R$, which in Euclidean space is clearly exponentially suppressed~\cite{Christ:2021guf}.

\begin{figure}
\centering
\includegraphics[width=0.8\linewidth]{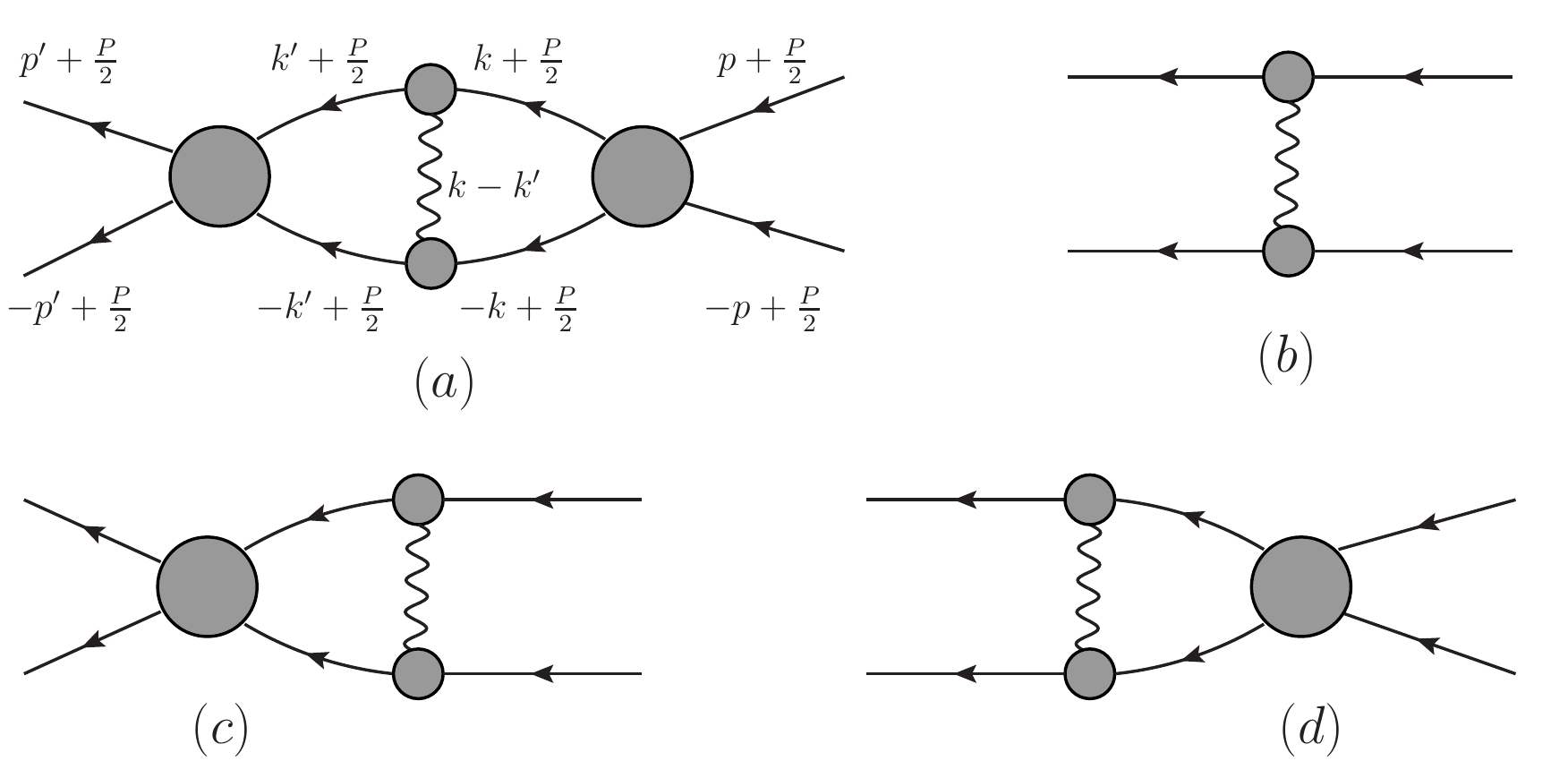}
\caption{The four classes of exchange diagrams depending on when the $\VCTC$ interaction occurs within the Lippman-Schwinger series. Evaluating all these diagrams is necessary to determining the correction to the phase shift from the long distance part of the Coulomb interaction.}
\label{fig:DWBA}
\end{figure}

The final set of diagrams $(c)$, called the exchange diagrams, can lead to power law corrections in $R^{-1}$ which we wish to determine. As shown in Fig.~\ref{fig:DWBA}, there exist 4 scenarios of how $\VCTC$ can be included into these types of diagrams. Though these diagrams generally contain the off shell scattering vertex and electromagnetic form factors, the long distance nature of $\VCTC$ restricts them to be the on shell quantities. These could in principle by calculated in a separate lattice QCD calculation or determined from either experiment or phenomenology. 

The contribution to $M_l$ to first order in $\VCTC$, for all four exchange diagrams, can be written as 
\begin{equation}
{\MCTC}_{,\,l} =\frac{1}{2l+1}\sum_{m=-l}^l \int \!\! \int d^4k' d^4k \, \Psi^\mathrm{out}_{lm}(k',P)^* \KCTC(k',k,P)\Psi^\mathrm{in}_{lm}(k,P). 
\label{eq:DWBA}
\end{equation}
where the momentum space scattering kernel for the $\VCTC$ interaction, after removing the momentum conserving delta function, is given by
\begin{eqnarray}
\KCTC(k',k,P) &=& -2\int\!\! \int d^4 y_2\, d^4 y_1 \int d^3 z \int \!\! \int d^4 x_2\, d^4 x_1 \;
e^{-i(k'+\frac{P}{2})y_2}e^{-i(-k'+\frac{P}{2})y_1} 
\label{eq:KCTC-MS} \\ 
&& \hskip 0.1 in
   \Bigl\langle \phi(y_2)\rho(\frac{z}{2})\phi^\dagger(x_2)\Bigr\rangle_{\mathrm{1PI}}   \VCTC(z)
   \Bigl\langle \phi(y_1)\rho(-\frac{z}{2})\phi^\dagger(x_1)\Bigr\rangle_{\mathrm{1PI}} e^{+i(k+\frac{P}{2})x_2}e^{i(-k+\frac{P}{2})x_1}. 
   \nonumber
\end{eqnarray}
and the wave function is decomposed between its non-interacting plane wave component, $\psi^0_{lm}$, and those in which the pions have scattered off each other, $\psi^\mathrm{in/out}_{lm}$,
\beq
\Psi^\mathrm{in/out}_{lm}(k,P) = \psi^0_{lm}(k,P) + \psi^\mathrm{in/out}_{lm}(k,P)\,.
\label{eq:wave-function-dcomp}
\eeq
The four possibilities of which component wavefunction is used for $\Psi^\mathrm{in/out}_{lm}$ which are shown in Fig.~\ref{fig:DWBA}.

The full analytical evaluation of Eq.~\eqref{eq:DWBA} is given in~\cite{Christ:2021guf}. Here we summarize the important final result of the $O(\alpha)$ correction to the phase shift from the $\VCTC$ interaction as
\begin{eqnarray}
\dCTC_l &=& \frac{p}{32\pi\omega_p} {\MCTC}_{,l} e^{-2\delta_l} \\
    &=& -p\omega_p\int_0^\infty w^2 dw 
\left\{\int\!\!\int d^3 r_2\, d^3 r_1 \overline{\rho}(r_2) \VCTC(\vec w - \vec r_2 + \vec r_1) \overline{\rho}(r_1) \right\} \nonumber \\
&& \hskip 0.5 in \cdot \Bigl[\cos{\delta_l}j_l(pw) + \sin(\delta_l)n_l(pw)\Bigr]^2 e^{-\mu w}  \,,
\label{eq:result}
\end{eqnarray}
where $j_l$ and $n_l$ are the spherical Bessel functions and $\bar{\rho}(r)$ is the Fourier transform of the electromagnetic form factor $F(q^2)$. The factor $\exp(-\mu w)$ serves to regulate the logarithmic singularity in the Coulomb interaction with a screening mass $\mu$. This singularity is the same as that which appears in the exact solutions of the Schr\"odinger and Dirac equations for the Coulomb potential and is caused by the long distance nature of the Coulomb interaction. This logarithmic contribution is a occurs for all partial waves in the same amount and will cancel in physical quantities such as $\eta_{+-}$ which is needed for CP violation in neutral kaon decays. 

\section{Conclusion}
Before working to calculate the isospin breaking corrections to the direct CP violation parameter $\epsilon'$, we have proposed a method to determine a piece of the isospin breaking contributions to $\pi^+\pi^+$ scattering, specifically the Coulomb potential. With QED in the Coulomb gauge, this problem is a well defined and separable piece of the full electromagnetic corrections to the scattering. A similar method will be necessary to find the Coulomb potential corrections to $K\to \pi\pi$ decays. 

In this method, the Coulomb interaction is broken into two distinct pieces, short distance and long distance. The short distance interaction, which occurs between charge operators separated by less than a scale $R$ that must be smaller than $L/2$, can be added directly into a standard lattice QCD calculation of the phase shift and L\"uscher quantization performed either non-perturbatively or the quantization condition can be expanded in perturbation theory. The long distance interaction can be analyzed in infinite volume analytically and the correction added to the result from L\"uscher quantization.

This long range correction can be determined up to certain limits. First, $R$ must be sufficiently large that the pions' interactions can be treated in a local $(\phi^\dagger\phi)^2$ theory. Second, the energy of the two pion system must remain under the four pion threshold. Within these limits, the correction to the phase shift can be determined up to terms exponentially suppressed in $R$. This feature, along with the exponentially suppressed finite volume corrections in L\"uscher quantization aside from those power law corrections from typically neglected $l\ge4$ contributions, means the QED corrected phase shift could be calculated to high accuracy.

There are two important steps moving forward before we can obtain the isospin breaking corrections to $\epsilon'$. First, the Coulomb interaction described here must be extended to the two channel system of $\pi^+\pi^-$ and $\pi^0\pi^0$ and to the $K\to \pi\pi$ decay amplitudes. Second, the transverse radiation, which was ignored, is required for the fully relativistic QED implementation to be correct. This piece is complicated by the effects of intermediate states with photons and the infrared divergences. 

\bibliographystyle{apsrev4-2.bst}
\bibliography{references}

\end{document}